\documentclass{aa}  

\usepackage{graphicx}
\graphicspath{{./}{Plots/}}
\usepackage{txfonts}
\usepackage{natbib}
\usepackage{color}
\usepackage{breakurl}
\usepackage{enumitem}
\usepackage{siunitx}
\usepackage{hyperref}
\hypersetup{colorlinks,citecolor=blue,urlcolor=blue,linkcolor=blue}
\usepackage{amsmath}
\usepackage{xcolor}

\usepackage[section]{placeins}
\usepackage{placeins}
\usepackage{float}
\let\Oldsection\section
\renewcommand{\section}{\FloatBarrier\Oldsection}
\let\Oldsubsection\subsection
\renewcommand{\subsection}{\FloatBarrier\Oldsubsection}
\let\Oldsubsubsection\subsubsection
\renewcommand{\subsubsection}{\FloatBarrier\Oldsubsubsection}

\newcommand{\tab}[1]{Table\,\ref{T:#1}}
\newcommand{\sect}[1]{Sect.\,\ref{S:#1}}

\newcommand{\fig}[1]{Fig.\,\ref{F:#1}}

\DeclareRobustCommand*{\unit}[1]{\def~{\,}\ensuremath{\mathrm{\,#1}}}

\usepackage{svg}
\newcommand{\orcid}[1]{\href{https://orcid.org/#1}{\includegraphics[width=10pt]{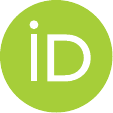}}}

\begin{document}

    \title{Coronal bright point statistics
\\ II. Magnetic polarities and mini loops}

  \author{I. Kraus \inst{1} \orcid{0000-0002-0451-811X},
          Ph.-A. Bourdin \inst{1, 2} \orcid{0000-0002-6793-601X},
          J. Zender \inst{3} \orcid{0000-0003-2728-3664},
          M. Bergmann \inst{3, 4} \orcid{0009-0005-5612-690X},
          A. Hanslmeier \inst{1}}

   \institute{Institute of Physics, University of Graz, Universit{\"a}tsplatz 5, 8010 Graz, Austria\\
              \email{Isabella.Kraus@uni-graz.at, Philippe.Bourdin@uni-graz.at}
         \and
             Space Research Institute, Austrian Academy of Sciences, Graz, Austria
         \and
             European Space Research and Technology Center, 2200 AG Noordwijk, Netherlands
          \and
             Iceye Oy, Maarintie 6, 02150 Espoo, Finland     }

   \date{Received 13 November 2024 / Accepted 8 January 2025}

 
  \abstract
   {The solar corona maintains temperatures of a million Kelvin or more. The plasma heating mechanisms responsible for these extreme temperatures are still unclear. Large regions of magnetic activity in the photosphere cause extreme ultraviolet (EUV) emission in the corona. Even smaller regions with bipolar and multipolar magnetic fields can generate coronal bright points (CBPs).}
   {We performed a statistical analysis of 346 CBPs. We used Solar Dynamics Observatory (SDO) images to track CBPs on a continuous basis. Therefore, we were able to collect a database of information on the CPB's lifetime, shape, polarity, flux emergence, and merging behavior, as well as their magnetic evolution, using the SDO Helioseismic and Magnetic Imager (SDO-HMI) instrument.}
   {We searched the SDO data archive for the longest continuous interval of uninterrupted observations in 2015. The longest such interval contains 12 consecutive days of full-disk images from the EUV channels of the SDO-AIA instrument. To analyze the properties of CBPs, we employed an automated tracking algorithm to follow the evolution of the CBPs. Furthermore, we manually checked the shape, underlying magnetic polarities, and merging behavior of each CBP.}
   {We provide statistics on the magnetic polarity, emergence, and merging of CBPs. We established a relationship between the CBP's merging behavior and both its shape and magnetic polarities. Brighter CBPs are visible in all SDO-AIA channels and exhibit strong radiative energy losses. The category of CBPs with a bipolar field has the highest probability of being emissive in all SDO-AIA channels. The majority of CBPs have two opposite polarities below them.}
   {The merging of two CBPs is an unusual phenomenon that is related to complex multipolar magnetic regions. Moreover, loop-shaped CBPs usually appear above bipolar fields. Faint CBPs have shorter lifetimes and are less likely to merge with another CBP.}


\authorrunning{I. Kraus et al.}

   \maketitle
%

\section{Introduction}


Coronal bright points (CBPs) are features on the Sun that involve hot plasma in the lower corona. The thermal energy of a CBP is mostly generated in the transition region and in the lower part of the corona \citep{2014AN....335.1037K,2014AstL...40..510M}. The first observations of CBPs were made with X-ray telescopes on board sounding-rocket flights \citep{1973SoPh...32...81V}.
Overall, CBPs are small bright structures in the corona \citep{1979SoPh...63..119S} and thousands of them can be seen all over the solar disk, simultaneously caused by magnetic perturbations \citep{2016Ap&SS.361..301L}.
The mean temperature of a CBP is about $1.3\unit{MK}$, but can rise to $3.4\unit{MK}$ \citep{2011A&A...526A..78K}.
The typical lifetimes of average CBPs are between several minutes to days and their size is smaller than 60\unit{arcsec} \citep{1973ApJ...185L..47V,2001SoPh..198..347Z}. They tend to appear above small bipolar magnetic-field regions \citep{2019LRSP...16....2M} and some look more complex and resemble tiny ARs \citep{2014AN....335.1037K,2014AstL...40..510M}.

The main polarities below the CBP have a total unsigned flux of about $10^{18}\unit{Mx}$ in the photosphere \citep{2013SoPh..286..125C}, while in the corona, a horizontal flux density of about $3-6\unit{G}$ has been suggested \citep{2001ApJ...553..429L}. The photospheric unsigned flux of the whole region below the CBP is approximately $2-3 \times 10^{19}\unit{Mx}$ \citep{1976SoPh...50..311G}.
The occurrence of CBPs is associated with magnetic reconnection forming small loops in about two-thirds of the cases, while the other third seems associated with emerging of flux tubes \citep{2001ApJ...553..429L}. Magnetic reconnection of bipolar fields may cause the generation of CBPs \citep{1976SoPh...50..311G,2007PASJ...59S.735K,2012Ap&SS.341..215L}.
\cite{2003ApJ...589.1062H} extended their study on CBP statistics with the eccentricity as an additional shape criterion. The topology of CBPs can be categorized as a simply point-like (roundish), loop-like, or complex shape.
There is no correlation between the number of CBPs and the solar cycle \citep{2002ApJ...564.1042S,2003ApJ...589.1062H,2011A&A...526A..78K}.

The magnetic energy released within one CBP is small, but with about 1000 CBPs visible on the solar disk simultaneously, a more substantial contribution to coronal heating is possible. However, above the quiet-sun area, about 5.2\% of the EUV emission stems from CBPs, but they only cover 1.4\% of the quiet-sun area \citep{2001SoPh..198..347Z}. The amount of energy released in one CBP is roughly $E_{CBP} = 10^{16}-10^{17}\unit{W}$ \citep{2019LRSP...16....2M}. On the whole sphere, there are about 2500 CBPs simultaneously above quiet Sun area, which we assume to cover 90\% of the solar surface. The global energy loss from CBPs, and hence their required heating, is about
\begin{equation}
P_{tot} = \frac{E_{CBP} \times 2500}{4 \pi ~ 700^2\unit{Mm^2} \times 0.9} = 4.5 - 45\unit{W/m^2} .
\end{equation}
The total required heating of the quiet Sun corona is about 100\unit{W/m^2}.
Based on these estimates, it has been determined that CBPs contribute about $4.5-45\%$ of the total quiet Sun coronal heating.

Throughout a CBP's lifecycle there are multiple flaring events possible \citep{2014AN....335.1037K,2014AstL...40..510M}. CBPs are uniformly distributed \citep{2014AN....335.1037K} in  quiet-Sun and coronal hole regions \citep{1974ApJ...189L..93G,1976SoPh...49...79G}. More details on CBPs are available in the review article of \cite{2019LRSP...16....2M}.

In the present work, we use the same CBP detection and tracking method as in \cite{2023A&A...678A.184K}. We also study additional properties, such as the magnetic polarity and merging of CBPs.


\section{Methods}
\label{S:methods}

We used Solar Dynamics Observatory (SDO) level-1 data with a spatial resolution of 0.6 arcsec and $4096^2$ pixels \citep{2012SoPh..275....3P,2012SoPh..275...17L,2012SoPh..275...41B}. The database consists of 12 consecutive days (13-24 August 2015). This was the longest available dataset in 2015 with continuous coverage. A Delta-V maneuver on 12th August 2015 and an Earth eclipse beginning at 25th of August 2015 were the causes of the limitation of the study. The reason was its circular geosynchronous orbit, where other transits and a Delta-V maneuver for orbit maintenance were possible. We used ``aia prep'' from the SolarSoft package to process images into level 1.5 data products. All images were co-rotated and co-aligned between the Atmospheric Imaging Assembly (AIA) and the Helioseismic and Magnetic Imager (HMI) \citep{2012SoPh..275..207S,2012SoPh..275..229S}.

We employed the SPoCA algorithm to automatically segment the solar disk in active regions (ARs), coronal holes (CHs), and the quiet sun (QS). The algorithm uses 171 and 193 channel images from the AIA instrument \citep{2014A&A...561A..29V,2014A&A...561A...9K}. Active regions are identified with the information from the HMI line-of-sight magnetograms \citep{2017A&A...605A..41Z}. The CBP identification is based on image morphological operators \citep{Haralick1987} applied to the high-resolution AIA 193 channel images \citep{2021SoPh..296..138V}.

\begin{figure}[!tbp]
\centering
\includegraphics[width=8.8cm]{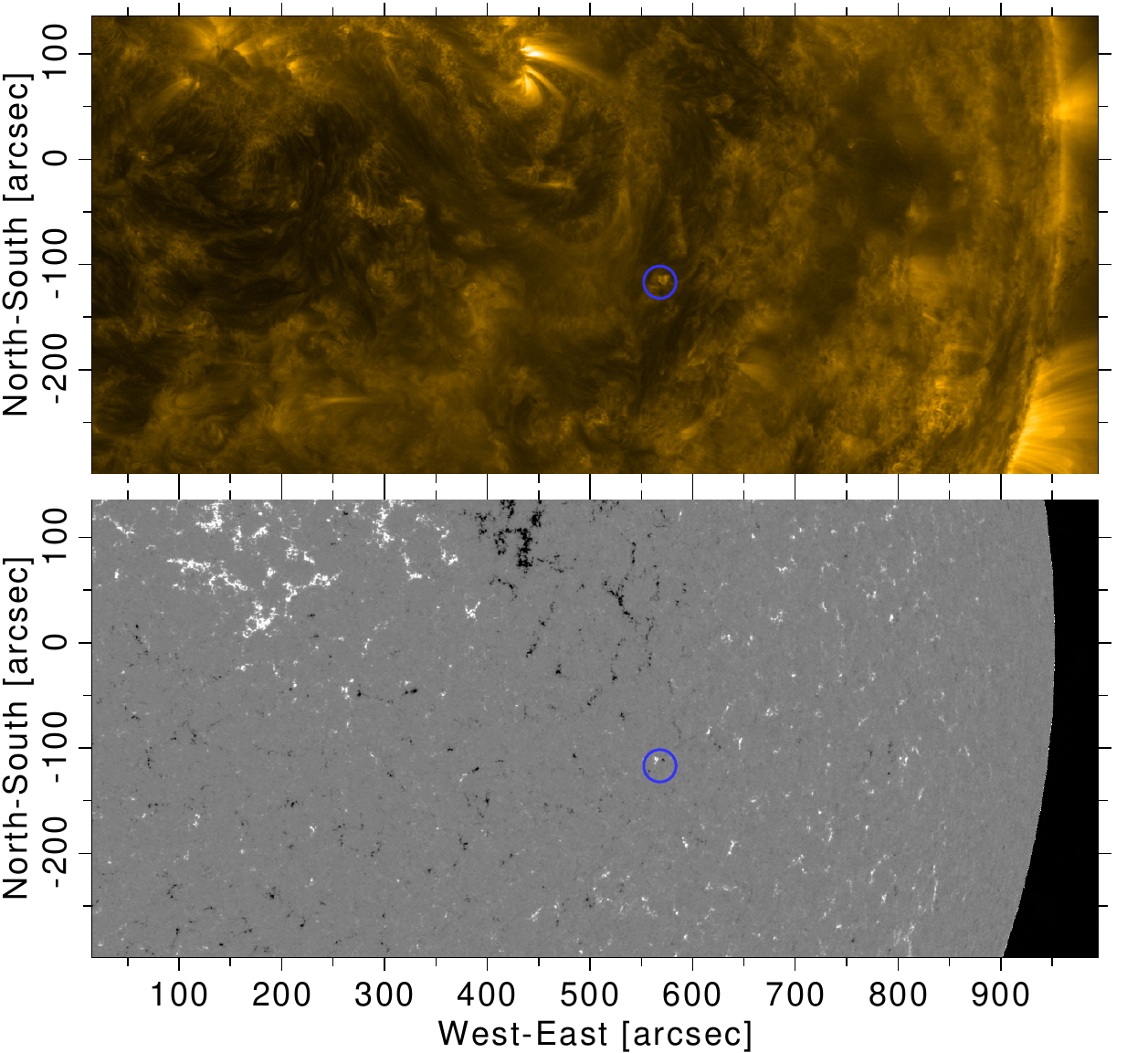}
\caption{EUV image of the solar disk observed by the SDO-AIA instrument in the 171\unit{\AA} channel (upper panel) and line-of-sight magnetic field from the SDO-HMI instrument (lower panel).
The images were taken on 16 August 2015 at 11:35:08UT.
The blue circle marks one sample CBP as identified by our tracking algorithm, see also the same CBP in \fig{Figure_2_4_2_Overview_Polarity}. The coordinates are provided in helioprojective format, with solar north oriented upwards. A video illustrating the evolution of this CBP can be accessed online (\href{https://zenodo.org/record/11370091}{https://zenodo.org/record/11370091}).}
\label{F:BP-869_2015-08-16_11_35_08_JHV}
\end{figure}

The automatic identification of a CBP is based on a set of selection criteria. These criteria include that the lifetime of a CBP is at least 60 min \citep{2015ApJ...807..175A}, the size is between 120 and 1500 arcsec \citep{2014ApJ...784..134R,2010SoPh..262..321S}, the shape has an eccentricity smaller than 2.5 \citep{2002A&A...392..329B,1979SoPh...63..119S}, the distance to the nearest AR is at least 180 arcsec \citep{2015ApJ...807..175A}, the distance from the limb is at least 5\% of the solar radius \citep{2015ApJ...807..175A}, and the intensity is significantly brighter than the background intensity in the AIA 171 channel \citep{2010A&A...516A..50S,2012A&A...538A..50S}. In \fig{BP-869_2015-08-16_11_35_08_JHV}, we show an example CBP and in our previous study we used the same criteria \citep{2023A&A...678A.184K}.

Our tracking method takes into account the latitude-dependent differential rotation of the Sun. The EUV flux of all CBPs is normalized with the maximum EUV flux of one reference CBP to make their images comparable. The CBP's EUV flux rises consistently across different wavelength channels. All identified CBPs are marked with contour lines and the frames are stacked with respect to the differential rotation. The appearance and disappearance of CBPs is taken from human-eye inspection of the 171 channel images, which determines the lifetime of each CBP.

In a previous study, we discussed also the EUV flux evolution of a single CBP and general CBP properties such as lifetime, shapes, and coronal co-rotation \cite{2023A&A...678A.184K}.

\section{Results}
      \label{S:3_Results}      

From the automatic CBP identification and tracking algorithm, we obtained sequences of the helioprojective coordinates of CBPs.
We did not use CBPs with an incomplete coverage due to tracking issues in our analysis. The remaining 346 CBPs were covered in time from before their appearance to after their disappearance. This allowed us to determine the lifetime of each CBP. The algorithm also draws contour lines of the CBP on each observed image, including the SDO-HMI magnetograms. Therefore, we were able to identify the magnetic patches below the CBP. Some examples are given in \fig{Figure_2_4_2_Overview_Polarity}.

\subsection{Flux emergence}
  \label{S:fluxemergence}
  
When a CBP is identified, the next step is to check back in time and determine what magnetic fields pre-existed at this location.
In only 3\% of the cases, we found no significant polarities below the later CBP area.
Instead, these cases have only random magnetic fluctuations in the photosphere, which is why we call these cases ``noisy'', as discussed in \sect{polarity}.
Noisy magnetic fields lead to very faint and short-lived CBPs that are usually visible only in the SDO-AIA 171 wavelength channel, which is explained by the fact that the 171 channel is the most sensitive to cool plasma.
Therefore, we suggest that those faint CBPs are not magnetically connected to strong polarities in the photosphere.

We also checked the magnetic evolution before the appearance of the CBP and we find that for all other 97\% of the cases, at least one magnetic polarity pre-exists and, in particular, the magnetic flux grows before the CBP appears.
This result shows that the CBP appearance is strongly connected with flux emergence.

\subsection{Magnetic evolution}
  \label{S:magneticevolution}

In this section we investigate the changes of CBPs only during the main phase of their lifetime.
This means that we do not discuss the very early and late evolution of the CBP.

\begin{table}[!tbp]
\centering
\caption{Events during the lifetime of a CBP, multiple options possible.}
\label{T:10_Interaction_Phase2}
\begin{tabular}{lr}
\hline \hline
evolution of CBP & \\
\hline
short separation in two parts (during main phase)          & 5.5\% \\  
change in shape category (during main phase)               & 0.3\% \\
change in size (during main phase)                         & 0.3\% \\
no changes in magnetic patches                             & 3.2\% \\
magnetic patches split (one polarity)                      & 4.3\% \\
magnetic patches split (both polarities)                   & 1.2\% \\
magnetic patches grow in size (one polarity)               & 1.5\% \\
magnetic patches grow in size (both polarities)            & 0.0\% \\
magnetic patches remain (one polarity)                     & 7.5\% \\
magnetic patches remain (both polarities)                  & 3.8\% \\
magnetic patches merge (one polarity)                      & 48.5\% \\
magnetic patches merge (both polarities)                   & 12.4\% \\
no merging of magnetic patches                             & 4.6\% \\
magnetic patches emerge (one polarity)                     & 2.9\% \\
magnetic patches emerge (both polarities)                  & 0.0\% \\  
no emergence of magnetic patches                           & 5.5\% \\
magnetic patches annihilate (fully)                        & 5.5\% \\
magnetic patches annihilate (partly)                       & 3.8\% \\
small-scale flux annihilation (one polarity remains)       & 27.4\% \\
small-scale flux annihilation (no polarity remains)        & 46.2\% \\
\hline
\end{tabular}
\end{table}

In 5.5\% of the cases we observe that a CBP may split into two parts for up to a few minutes.
After such a separation, the CBP merges again into one and remains one CBP.
We never observe CBPs that split and separate for longer time.

In \tab{10_Interaction_Phase2}, we present the fraction of CBPs with specific events during their main phase.
We analyzed the changes in the magnetic patches below the CBP area.
Emerging magnetic flux is identified as a flux concentration that appears in a quiet Sun region.
Annihilating flux patches would either fully disappear or some flux concentration would remain after the partly annihilation.
We find that all newly emerging magnetic flux either merges with the same polarity or annihilates with a pre-existing opposite polarity. The vast majority of emerging flux are tiny patches of small spatial scales that annihilate with a strong opposite polarity. As a result, either the pre-existing polarities may fully annihilate over time or one polarity remains after the CBP disappeared. Overall, 73.6\% of all CBPs fall into these two previous categories (see last two lines in \tab{10_Interaction_Phase2}).

If there are significant magnetic patches below a CBP, we find that these polarities may merge with the same type of polarity.
This may happen for either one magnetic polarity only or for both polarities, which together happens for 60.9\% of all CBPs; see the categories ``magnetic patches merge'' in \tab{10_Interaction_Phase2}.

During the main phase of the CBP's lifetime we find that the magnetic patches fully disappear in 51.7\% of the cases, either through full annihilation of opposite polarities or through continuous annihilation with small-scale flux. This can be seen in the row labeled ``magnetic patches annihilated fully'', along with the last line of \tab{10_Interaction_Phase2}.

For only 3.2\% of the CBPs we see no changes in their magnetic polarities at all.
These are the same CBPs that have also a noisy magnetogram without clear magnetic polarity (see also \fig{Polarity}).

As a consequence, almost all CBPs do not significantly change in their size or shape category during their main phase of their lifetime. In contrary, almost all CBPs do feature significant changes in the underlying magnetic polarities.

\subsection{Polarity}
  \label{S:polarity}

The HMI magnetograms indicate which magnetic polarities exist below the bright point. This could be only ``noisy'' data without a significant polarity in the direct vicinity of the CBP. If only one polarity is found, we call these ``unipolar''. Normally, there are two opposite polarities of roughly similar magnetic fluxes, forming a ``bipolar'' region. ``Multipolar'' regions feature both magnetic polarities with more than two magnetic patches. Such multipolar regions can either balance out from their signed magnetic fluxes or they are ``strongly unbalanced''.

\begin{figure}[!tbp]
\centering
\includegraphics[width=8cm]{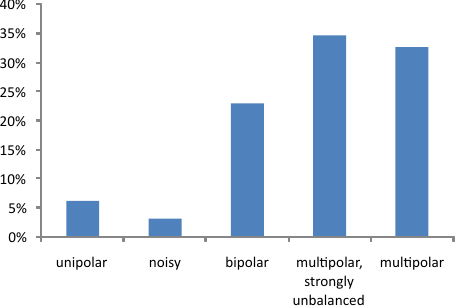}
\caption{Distribution of footpoint polarities of CBPs (see \sect{polarity}).}
\label{F:Polarity}
\end{figure}

In \fig{Polarity}, we display the distribution of magnetic polarities in the photosphere at the footpoints of CBPs. The vast majority of 91\% of the CBPs show both magnetic polarities at their footpoints. Only 6\% have only one magnetic polarity (``unipolar''), while another 3\% show no significant polarity (``noisy''). As field lines are usually connected on both ends to the photosphere, such CBPs with either unipolar or noisy magnetic field must have connectivity to regions located further away from the CBP. We find that ``multipolar'' regions usually have polarities of smaller area.

In the online material we show a video of a unipolar CBP with mainly negative polarity (2015-08-15 05:37:23 UT, \href{https://zenodo.org/record/11352890}{https://zenodo.org/record/11352890}), a unipolar CBP with mainly positive polarity (2015-08-14 10:21:39 UT, \href{https://zenodo.org/record/11280907}{https://zenodo.org/record/11280907}), a CBP with noisy magnetic fields (2015-08-15 17:04:23 UT, \href{https://zenodo.org/record/11371292}{https://zenodo.org/record/11371292}), a clearly bipolar CBP (2015-08-16 11:35:08 UT, \href{https://zenodo.org/record/11370091}{https://zenodo.org/record/11370091}), a multipolar CBP with a strongly unbalanced magnetic flux (2015-08-16 00:21:38 UT, \href{https://zenodo.org/record/11370230}{https://zenodo.org/record/11370230}), and a multipolar CBP with roughly balanced flux (2015-08-20 17:54:38 UT, \href{https://zenodo.org/record/11370380}{https://zenodo.org/record/11370380}).

\begin{figure}[!tbp]
\centering
\includegraphics[width=11cm]{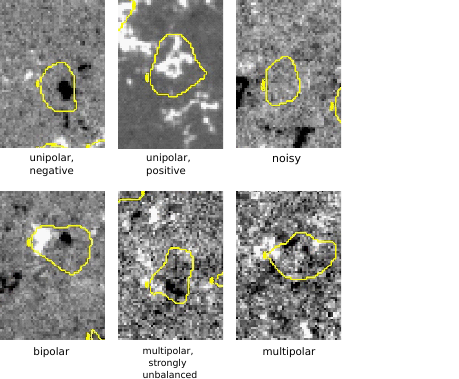}
\caption{Overview of CBP polarity classes, first row from left to right: unipolar negative (\href{https://zenodo.org/record/11352890}{https://zenodo.org/record/11352890}), unipolar positive (\href{https://zenodo.org/record/11280907}{https://zenodo.org/record/11280907}), noisy (\href{https://zenodo.org/record/11371292}{https://zenodo.org/record/11371292}); second row from left to right: bipolar (\href{https://zenodo.org/record/11370091}{https://zenodo.org/record/11370091}), multipolar strongly unbalanced (\href{https://zenodo.org/record/11370230}{https://zenodo.org/record/11370230}), multipolar (\href{https://zenodo.org/record/11370380}{https://zenodo.org/record/11370380}). The size of each panel is about $30 \times 42~\unit{arcsec}$ or $50 \times 70$ SDO pixels.}
\label{F:Figure_2_4_2_Overview_Polarity}
\end{figure}

We note that the color scale in the videos, as well as in \fig{Figure_2_4_2_Overview_Polarity}, is not the same because we adapted the coloring to improve the visibility for the polarities.

\subsection{Visibility versus polarity}
  \label{S:EUVemissivity}

\begin{figure}[!tbp]
\centering
\includegraphics[width=8cm]{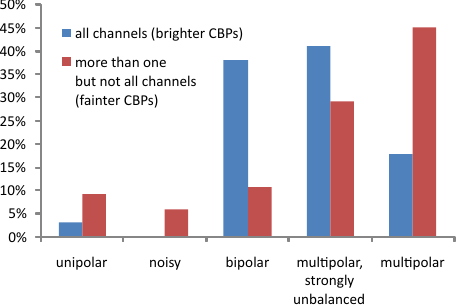}
\caption{Comparison of polarities for bright (blue) and faint (red) CBPs, where bright CBPs are visible in all AIA wavelength channels and faint CBPs are not (see \sect{EUVemissivity}). Each category of bright and faint CBPs is normalized to 100\%.}
\label{F:2_Polarity_Hot_vs_faint_CBP}
\end{figure}

About 46\% of our ensemble are bright CBPs that are visible in all AIA wavelength channels, while we call 54\% of the CBPs ``fainter'' because they are not visible in all AIA channels \citep{2023A&A...678A.184K}.
We compare the magnetic polarity of bright versus faint CBPs. From \fig{2_Polarity_Hot_vs_faint_CBP}, we see that bipolar CBPs are strongly biased to be bright. Furthermore, multipolar CBPs, with a strongly unbalanced magnetic flux at the footpoints have a tendency to be brighter than multipolar CBPs with a more balanced magnetic flux that have a strong tendency to be fainter.
Notably, we find no bright CBPs with a noisy magnetic field in \fig{2_Polarity_Hot_vs_faint_CBP}, while fainter CBPs have a significantly larger fraction. Similarly, about three quarters of the unipolar CBPs are faint, while about one quarter of them is bright.

\subsection{Shape versus polarity}
  \label{S:polarityvsshape}

In previous work we determined the typical shapes of CBPs \citep{2023A&A...678A.184K}; see \tab{Shape}.
We now compare the CBP shapes according to their magnetic polarities at their footpoints.

In the quiet-Sun magnetic network one would expect a similar behavior of unipolar CBPs, irrespective of the sign of their polarity. We confirm this is the case in our data, where any differences between positive and negative unipolar CBPs are fully within our statistical errors. Therefore, we treat all unipolar CBPs as a single group.

\begin{table}[!tbp]
\centering
\caption{Distribution of CBP shape classes from \cite{2023A&A...678A.184K}.}
\label{T:Shape}
\begin{tabular}{lr}
\hline \hline
shape category & fraction of CBPs \\
\hline
roundish/simple	    & 16\% \\
loop-like/curved	    & 49\% \\
complex		    & 35\% \\
\hline
\end{tabular}
\end{table}

\begin{figure}[!tbp]
\centering
\includegraphics[width=8.8cm]{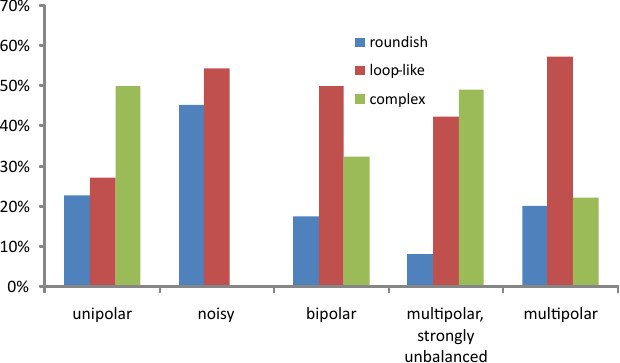}
\caption{Distribution of CBP polarities for different shapes (see \sect{polarityvsshape}). Each polarity category is normalized to 100\%.}
\label{F:Polarity_vs_Shape}
\end{figure}

From \fig{Polarity_vs_Shape} we find that CBPs without a clear polarity (noisy) have either a ``loop-like'' shape (55\%) or have ``roundish`` shapes (45\%).
These percentages are not significantly different due to the statistical error within the ensemble of noisy CBPs (see \tab{Shape}).
Without any clear magnetic polarity below a CBP we find no ``complex'' shapes.
This already gives a hint that the shape of a CBP is probably dominated by the topology of the magnetic field.
For the bipolar group of CBPs, we find their shape distribution is very similar to the distribution of all CBPs in \tab{Shape}.
It is only for the group of multipolar fields that we can see a strikingly larger fraction of loop-like shapes.
``Multipolar strongly unbalanced`'' CBPs have significantly more complex shapes than the full ensemble of all CBPs.

\subsection{Polarity versus lifetime}
  \label{S:lifetimevspolarity}

\begin{figure}[!tbp]
\centering
\includegraphics[width=8.8cm]{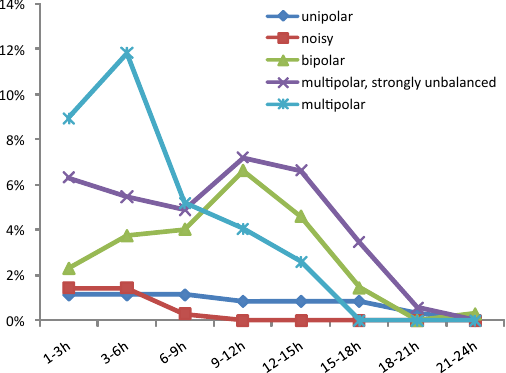}
\caption{Lifetimes of CBPs according to their polarity classes, binned in three-hours time intervals (see \sect{lifetimevspolarity}). The sum of all data points is normalized to 100\%.}

\label{F:6_Lifetime_vs_Polarity}
\end{figure}

We relate the different polarities below CBPs from \fig{Polarity} to their lifetimes.
From \fig{6_Lifetime_vs_Polarity}, we see that most of the CBPs have lifetimes up to 18 hours and only very few CBPs have longer lifetimes up to 24 hours.
For most of the CBPs with noisy polarities we find lifetimes of only up to 9 hours. Since there is no significant polarity below such CBPs, these shorter lifetimes are to be expected.
Bipolar CBPs show a significant peak of lifetimes from 6 to 15 hours.
Multipolar CBPs with strongly unbalanced flux have in general a very similar distribution of lifetime as the bipolar ones, but with a significant increase for lifetimes of 6 hours and less.
Multipolar CBPs with roughly balanced flux show significantly shorter lifetimes with the majority of 6 hours or less and a quick decay of the distribution for longer lifetimes.
The groups of bipolar and multipolar balanced regions do not show significant differences beyond statistical errors for lifetimes of 6 hours or longer.

\subsection{Merging CBPs}
  \label{S:mergingbrightpoints}

We speak of a merger of CBPs when two CBPs approach and become one indistinguishable object. In all of the merging CBPs cases, we find that after less than one hour, they separate again. We find the probability of one CBP to merge with another one is 36\%, while for mergers we count these CBPs only once. The majority of 64\% of the CBPs remain isolated. For an example of a CBP merger see online video (2015-08-16 05:22:24 UT, \href{https://zenodo.org/record/11370492}{https://zenodo.org/record/11370492}).

\subsection{Merging versus shape}
  \label{S:shapevsmergingbrightpoints}

\begin{figure}[!tbp]
\centering
\includegraphics[width=7.5cm]{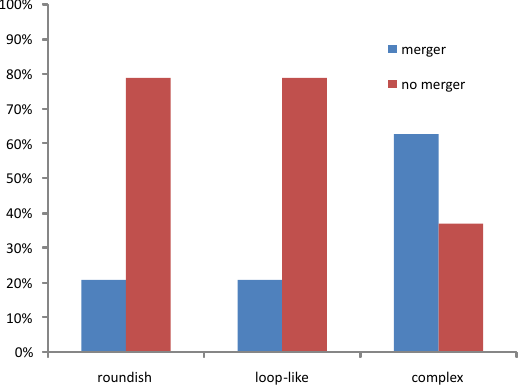}
\caption{Fraction of merging CBPs with respect to their shape category (see \sect{shapevsmergingbrightpoints}). Each shape category is normalized to 100\%.}
\label{F:2_Shape_vs_Merging}
\end{figure}

In \fig{2_Shape_vs_Merging}, we display the correlation between CBP shape category and the probability of mergers. Our results demonstrate that CBPs with roundish and loop-like shapes have lower probabilities of merging with another CBP, which is around 20\%. In contrast to that, we find CBPs with complex shapes have a significantly higher probability of merging.

\subsection{Merging versus lifetime}
  \label{S:lifetimevsmergingbrightpoints}

\begin{figure}[!tbp]
\centering
\includegraphics[width=8.8cm]{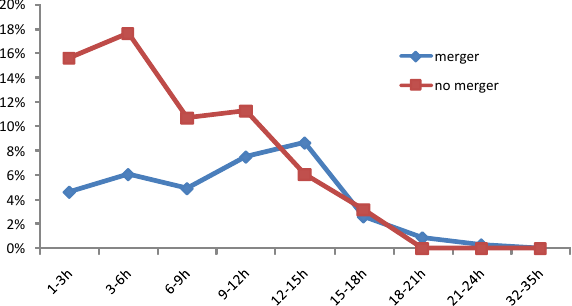}
\caption{Distribution of lifetimes of CBPs for the groups of merging (blue) and non-merging (red) CBPs (see \sect{lifetimevsmergingbrightpoints}). The sum of all data points is normalized to 100\%.}
\label{F:4_Lifetime_vs_Merging}
\end{figure}

We now investigate how the lifetime and the occurrence of CBP mergers are related.
In \fig{4_Lifetime_vs_Merging}, we show the lifetime distribution of CBPs in two groups: merging CBPs (blue) and non-merging ones (red).
For lifetime up to 9 hours we find a significant larger fraction of CBPs that do not merge with other ones.
For longer lifetimes, the differences are not statistically significant.

\subsection{Merging versus polarity}
  \label{S:polarityvsmergingbrightpoints}

\begin{figure}[!tbp]
\centering
\includegraphics[width=8.8cm]{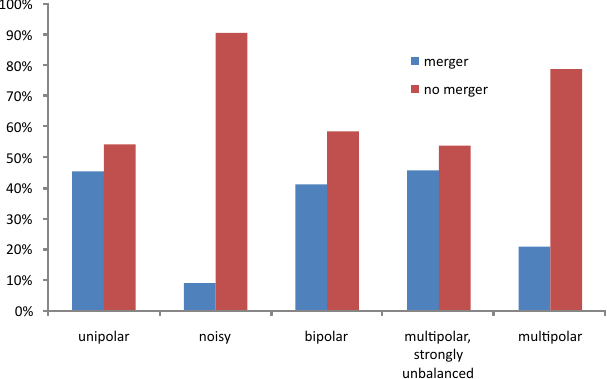}
\caption{Merging behavior of CBPs for multiple classes of magnetic polarities (see \sect{polarityvsmergingbrightpoints}). Each polarity class is normalized to 100\%.}
\label{F:3_Polarity_vs_Merging}
\end{figure}

The global average probability for CBP mergers is 36\%. From \fig{3_Polarity_vs_Merging} we see that the probability of merging in the unipolar and ``multipolar strongly unbalanced'' categories is significantly higher with about 45\%. In the bipolar category, CBPs merge with a probability of more than 40\%, which is slightly higher than for the global average. In the noisy category (showing no clear polarities), we find only about 10\% of CBPs merge. The multipolar category features a lower probability of merging in about 20\% of the cases.

\subsection{Mini loops}
  \label{S:mini-loops}

There is ongoing discussion if a CBP is a very small coronal loop or if a CBP is at least associated with coronal loops \citep{2019LRSP...16....2M}.
We characterize ``mini loops'' as elongated and curved shapes visible in EUV emission, in particular the AIA 171 channel.
Most probably, such mini loops are composed of hot material that expands along the magnetic field and radiates in EUV.

We identify mini loops as loop-like, elongated and often curved structures originating from a CBP.
We find that about 75\% of all CBPs in our study feature mini loops for most of their lifetime and the mini loops are visible in three wavelength channels simultaneously; see AIA 171, 193, and 211 channels in \tab{Table_3_6_1_Fluxtube_probability_channel}.
The typical length of these mini loops is about $15-20\unit{Mm}$, which implies the loop has a height of about $7.5-10\unit{Mm}$.

The presence of a mini loop likely indicates the magnetic connectivity between a CBP and a nearby region on the Sun or between two CBPs.
Whereas the absence of mini loops likely indicates that the CBP is isolated or magnetically connected to a region far away from the CBP, which usually implies a nearly vertical magnetic field configuration.

Only about 15\% of all CBPs feature no mini loops at all.
Since 46\% of all CBPs are visible in all AIA channels \citep{2023A&A...678A.184K}, but almost no mini loops are visible in these channels (\tab{Table_3_6_1_Fluxtube_probability_channel}), it is likely that mini loops have cooler temperatures as the CBPs themselves.
In the accompanying online video, we show a well visible mini loop connected to a CBP (2015-08-16 16:19:23 UT, \href{https://zenodo.org/record/11370642}{https://zenodo.org/record/11370642}).

\begin{table}[!tbp]
\centering
\caption{Visibility of mini loops at CBPs, multiple options possible.}
\label{T:Table_3_6_1_Fluxtube_probability_channel}
\begin{tabular}{rr}
\hline \hline
AIA channel & fraction of CBPs \\
\hline
304 \unit{\AA} &  36\%\\
131 \unit{\AA} &   5\%\\
171 \unit{\AA} &  75\%\\
193 \unit{\AA} &  77\%\\
211 \unit{\AA} &  73\%\\
335 \unit{\AA} &   1\%\\
94 \unit{\AA}  &   1\%\\
\hline
\end{tabular}
\end{table}


\section{Discussion}
      \label{S:Discussion}
The emergence of a CBP is likely attributed to magnetic energy release during magnetic reconnection \citep{2003A&A...398..775M}. An increase in photospheric magnetic flux results in enhanced energy production in the lower corona, subsequently leading to intensified EUV emission or larger CBPs \citep{2013SoPh..286..125C}. This is explained by the fact that higher magnetic flux densities lead to stronger Poynting fluxes and more magnetic energy is subsequently provided to the corona \citep{2014PASJ...66S...7B,2015A&A...580A..72B}. Scaling laws on coronal heat input also have the magnetic flux density as contributing parameter with varying exponents \citep{2016A&A...589A..86B}. Similarly, particle acceleration due to electric fields in the corona scales also with the magnetic flux density \citep{2016A&A...587A...4T}.

We found that faint CBPs usually have only weak magnetic fields that resemble a noisy magnetogram.
This implies the magnetic connectivity is not likely to be directed toward the photosphere, but instead to a region farther away from the CBP.
Such field configurations should have a more horizontal magnetic-field vector and are likely the result from previous magnetic reconnection, leading to a heating near the apex of a mini loop, similar to what was described by \cite{2001ApJ...553..429L}.

In the typical evolution of a CBP, we see first some magnetic patches in the photospheric field before the CBP appears in EUV wavelengths. After some time, the CBPs become EUV emissive because magnetic disturbances from the photosphere need the Alfv{\'e}n travel time of about 3--20 min to reach the lower corona. It is only after this time that the corona may dissipate magnetic energy injected from the photosphere and heat the coronal plasma to become EUV emissive.

After the CBP became large and bright enough, the tracking algorithm is stable during the whole lifetime. Near the end of the lifetime, the main magnetic polarities approach each other, merge, and ultimately become annihilated. After either one or both magnetic polarities have disappeared or annihilated, the CBP will shortly later start to fade out until it completely disappears in all AIA channels. In the online video we show the AIA wavelength channels along with the HMI magnetograms of a typical CBP lifetime (2015-08-21 09:39:38 UT, \href{https://zenodo.org/record/11370968}{https://zenodo.org/record/11370968}).

\section{Conclusions}
      \label{S:Conclusions} 

In our ensemble of CBPs, 91\% of them have been found to form between two opposite magnetic polarities that are distributed in either two or more individual magnetic patches (see \sect{polarity}).
Those magnetic patches are stable over most of the main phase of the lifetime of the CBPs.
The EUV intensity maximum is typically located between the two polarities.
In about 73.6\% of the cases we see either full or partial annihilation of the polarities shortly before the disappearance of the CBP (see \sect{magneticevolution}).
We conclude that the magnetic polarities below the CBP are therefore strongly related to the heating process that maintains the heat input for the CBP, whereas it radiates strongly in EUV.

Most of our CBPs form above opposite magnetic polarities, whereas only 9\% of the CBPs form above unipolar or noisy magnetic fields.
For multipolar regions with a roughly balanced magnetic flux, as well as for bipolar regions, we find a significantly larger fraction of loop-like shapes among CBPs (see \sect{polarityvsshape}).
This suggests that the magnetic-field configuration connects mainly between the underlying opposite polarities and, hence, the field follows roughly a semi-circular shape.
Instead, if there is a multipolar field with a strongly unbalanced flux below a CBP, this results in a more diverse field topology, hence, we would expect more complex shapes for CBPs. This is confirmed on the basis of our results, shown in \fig{Polarity_vs_Shape}.

When we look at the co-appearance of mini loops, we find that about 75\% of all CBPs are connected to a EUV-emissive structure resembling a small-scale coronal loop.
This could indicate either a magnetic connectivity to another nearby CBPs or a strongly asymmetric heating within a small-scale coronal loop.
Mini loops reach to a height of about $7.5-10\unit{Mm}$, which is in the lower corona.
We suggest to observe mini loops mainly in the AIA 171, 193, and 211 channels because, in our ensemble, most of the mini loops are seen simultaneously in exactly these channels (see \tab{Table_3_6_1_Fluxtube_probability_channel}).

Almost all analyzed CBPs that are visible in all AIA channels do appear above regions with opposite magnetic polarities (see blue bars in \fig{2_Polarity_Hot_vs_faint_CBP}).
Obtaining a good visibility for CBPs in all AIA channels requires higher coronal temperatures and, hence, a greater amount of heating than for fainter ones.
We argue that the heating is caused by dissipation of magnetic energy, namely, a weak magnetic reconnection.

Unipolar CBPs have a significantly shorter lifetime.
This can be understood because unipolar field would typically lead to more vertical field configurations with less curvature and a much smaller possibility for magnetic reconnection and dissipation of generated currents. Therefore, CBPs above unipolar field are less bright in EUV.

In summary, CBP formation and heating mechanisms are not primarily driven by the transport of hot plasma to increase the thermal energy in the corona, whereas magnetic energy dissipation is a plausible heating mechanism for CBPs.

\begin{acknowledgements}
SDO Data supplied courtesy of the SDO/HMI and SDO/AIA consortia.
This research was funded in whole or in part by the Austrian Science Fund (FWF) [10.55776/P32958] and [10.55776/P37265].
\end{acknowledgements}

\newpage

\bibliographystyle{aa}
\bibliography{Literatur_Isabella_Kraus}

\end{document}